\documentclass[11pt]{article}
\usepackage{epsfig}

\def\la{\langle}\def\ra{\rangle}
\def\be{\begin{eqnarray}}
\def\ee{\end{eqnarray}}
\def\lsim{\mathrel{\rlap{\lower3pt\hbox{\hskip1pt$\sim$}}
     \raise1pt\hbox{$<$}}} 
\def\gsim{\mathrel{\rlap{\lower3pt\hbox{\hskip1pt$\sim$}}
     \raise1pt\hbox{$>$}}} 

\def\pl{Phys. Lett.}
\def\bi{\bibitem}
\def\L{{\cal L}}
\begin{document}
\begin{titlepage}

\hfill {\today: hep-ph/}

\begin{center}
\ \\
{\Large \bf  The Role of the Dilaton in  Dense Skyrmion Matter}
\\
\vspace{.30cm}

Byung-Yoon Park$^{a}$, Mannque Rho$^{b}$
\\ and Vicente Vento$^{c}$

\vskip 0.20cm

{(a) \it Department of Physics,
Chungnam National University, Daejon 305-764, Korea}\\
({\small E-mail: bypark@cnu.ac.kr})

{(b) \it Service de Physique Th\'eorique, CE Saclay}\\
{\it 91191 Gif-sur-Yvette, France}\\
({\small E-mail: rho@spht.saclay.cea.fr})

{(c) \it Departament de Fisica Te\`orica and Institut de
F\'{\i}sica
Corpuscular}\\
{\it Universitat de Val\`encia and Consejo Superior
de Investigaciones Cient\'{\i}ficas}\\
{\it E-46100 Burjassot (Val\`encia), Spain} \\ ({\small E-mail:
Vicente.Vento@uv.es})

\end{center}
\vskip 0.3cm

\centerline{\bf Abstract}
In this note, we report on a remarkable
and surprising interplay between the $\omega$  meson  and the
dilaton $\chi$  in the structure of a single skyrmion as well as in
the phase structure of dense skyrmion matter which may have a
potentially important consequence on the properties of compact
stars. In our continuing effort to understand hadronic matter at
high density, we have developed a unified field theoretic formalism
for dense skyrmion matter using a single Lagrangian to describe
simultaneously both matter and meson fluctuations and studied {\it
in-medium} properties of hadrons. The effective theory used is the
Skyrme model Lagrangian gauged with the vector mesons $\rho$ and
$\omega$, implemented with the dilaton field that describes the
spontaneously broken scale symmetry of QCD,  in a form consistent
with the symmetries of QCD and our expectations regarding the high
density limit. We analyze the restoration of scale invariance and
chiral symmetry as the density of the system increases. In order to
preserve the restoration of scale symmetry and chiral symmetry,
signalled in our case by the vanishing of the expectation value of
the dilaton, {\em and} to be consistent with the ``vector
manifestation" of hidden local symmetry, a density dependent
$\omega$ coupling is introduced. We uncover  the crucial role played
by both the dilaton and the $\omega$ meson in the phase structure of
dense medium and discover how two different phase transition regimes
arise as we ``dial" the dilaton mass.

\vskip 0.5cm

\vskip 0.3cm \leftline{Pacs: 12.39-x, 13.60.Hb, 14.65-q, 14.70Dj}
\leftline{Keywords: skyrmion, dilaton, vector mesons, dense matter}

\end{titlepage}
\section{Introduction}
Hadronic matter at high density is presently poorly understood, and
the issue of the equation of state (EOS) in the density regime
appropriate for the interior of compact stars remains a wide open
problem.  In the absence of model-independent theoretical tools such
as lattice and of experimental guidance for dense hadronic matter,
it is difficult to assess the reliability -- perhaps even relevance
-- of the plethora of scenarios predicted by a variety of models for
compact stars available in the literature. It is in this context
that the skyrmion approach to dense matter anchored on large $N_c$
QCD, with density effects simulated on crystal lattice, was put
forward in  a series of papers~\cite{byp1,byp2,vector} as a
potentially promising and consistent approach.

Motivated by a recent development on hidden local symmetry (HLS)
approach to low-energy effective field theory for
hadrons~\cite{HY:HLS} which gave -- at least in the chiral limit --
an elegant and unambiguous  prediction of the behavior of
light-quark hadrons at high temperature and/or at high
density~\footnote{In this note, we will be mainly considering
density effects. However many of the arguments used for density hold
also for temperature.}, both the dilaton filed $\chi$ and the vector
meson fields $\rho$ and $\omega$ were incorporated into the Skyrme
Lagrangian in \cite{vector} to construct dense skyrmion matter on
crystal lattice. In HLS theory of \cite{HY:HLS}, combining local
gauge invariance of the flavor gauge fields $\rho$ and $\omega$ and
matching of the low-energy effective theory to QCD at a suitable
matching scale uncovered a unique fixed point -- called ``vector
manifestation (VM)" fixed point -- to which dense  matter flows as
the density increases toward the critical chiral phase transition
point. The unambiguous prediction was that the hidden gauge coupling
constant $g$ and the physical pion decay constant $f_\pi$ would
vanish at that point, with the consequence that the vector meson
mass would vanish at the chiral transition. This meant  that the
vector mesons would be essential degrees freedom in many-nucleon
dynamics under extreme conditions.  Now the Lagrangian used in
\cite{vector} is a gauge fixed one -- at unitary gauge -- and hence
lacks the intrinsic dependence that represents matching to QCD
present in hidden local gauge invariant theory. What the scalar
dilaton does is,  as first described in \cite{BR91},  to simulate,
albeit approximately,  in the gauged Skyrme Lagrangian the intrinsic
dependence that figures importantly  in HLS theory.

What was found in \cite{vector} was disturbing and, at the same
time, highly interesting. In the absence of the vector mesons,  the
presence of the dilaton provides a reasonable scenario for chiral
restoration~\cite{byp2}. However in the presence of the vector
mesons, in particular, the $\omega$ meson, while chiral symmetry is
restored at some density with the order parameter $\la\sigma\ra\sim
\la\bar{q}q\ra\rightarrow 0$ -- which is required by symmetry on
crystal lattice~\cite{half-skyrmions}, the pion decay constant
$f_\pi$ whichi is tagged to the expectation value of the $\chi$
field $\la\chi\ra^*$ (where the asterisk stands for medium) does not
vanish at the point where the condensate is zero. This aspect is
related to the fact that $\omega$ mass cannot vanish in the model --
in fact, it increases at higher density -- which is at odds with the
vector manifestation of HLS theory. This means a strong repulsion
developing as density increases, an aspect which is qualitatively
important for nuclear physics.

The objective of this paper is to precisely identify the problem
involved and suggest possible resolutions to the problem.

The content of this paper is as follows. In Section 2, we pinpoint
what the problem is. In Section 3 is given our proposed solution to
the problem. In Section 4, the properties of a single skyrmion are
studied. We pay special attention to the role of the dilaton field
and the spatial structure of the realization of scale and chiral
symmetry within the skyrmion. Section 5 is devoted to the study of
the  phase transitions in dense skyrmion matter and the fundamental
role of the dilaton field in their realizations. Some concluding
remarks are given in Section 6.

\section{The Problem}
In order to clarify  the problem involved, we first discuss the
Lagrangian used in \cite{vector}: \be {\cal L}=\L_n +
\L_{an}\label{lag} \ee where
\begin{eqnarray}
{\cal L}_n &=& \frac{f_\pi^2}{4} \left(\frac{\chi}{f_\chi}\right)^2
\mbox{Tr}(\partial_\mu U^\dagger \partial^\mu U) + \frac{f_\pi^2
m_\pi^2}{4} \left(\frac{\chi}{f_\chi}\right)^3
    \mbox{Tr}(U+U^\dagger-2)
\nonumber\\
&&
-\frac{f_\pi^2}{4} a \left(\frac{\chi}{f_\chi}\right)^2
 \mbox{Tr}[\ell_\mu + r_\mu + i(g/2)
( \vec{\tau}\cdot\vec{\rho}_\mu + \omega_\mu)]^2
-\textstyle \frac{1}{4} \displaystyle
\vec{\rho}_{\mu\nu} \cdot \vec{\rho}^{\mu\nu}
-\textstyle \frac{1}{4}  \omega_{\mu\nu} \omega^{\mu\nu}
\nonumber\\
&&
+\textstyle\frac{1}{2} \partial_\mu \chi \partial^\mu \chi
-\displaystyle \frac{m_\chi^2 f_\chi^2}{4} \left[ (\chi/f_\chi)^4
(\mbox{ln}(\chi/f_\chi)-\textstyle\frac14) + \frac14 \right],
\label{lag-n}\\
\L_{an} &=& \textstyle\frac{3}{2} g  \omega_\mu B^\mu\label{lag-an}
\end{eqnarray}
where

$$
U =\exp(i\vec{\tau}\cdot\vec{\pi}/f_\pi) \equiv \xi^2,
\eqno(\mbox{\ref{lag}.a})$$
$$
\ell_\mu = \xi^\dagger \partial_\mu \xi, \mbox { and }
r_\mu = \xi \partial_\mu \xi^\dagger,
\eqno(\mbox{\ref{lag}.b})$$
$$
\vec{\rho}_{\mu\nu} = \partial_\mu \vec{\rho}_\nu
- \partial_\nu \vec{\rho}_\mu + g \vec{\rho}_\mu \times \vec{\rho}_\nu,
\eqno(\mbox{\ref{lag}.c})$$
$$
\omega_{\mu\nu}=\partial_\mu\omega_\nu-\partial_\nu\omega_\mu,
\eqno(\mbox{\ref{lag}.d})$$
$$
B^\mu =  \frac{1}{24\pi^2} \varepsilon^{\mu\nu\alpha\beta}
\mbox{Tr}(U^\dagger\partial_\nu U U^\dagger\partial_\alpha U
U^\dagger\partial_\beta U).
\eqno(\mbox{\ref{lag}.e})$$
The Lagrangian (\ref{lag-n}) is the normal part of the flavor $U(2)$
chiral Lagrangian that includes the vector mesons $\rho$ and
$\omega$ and the dilaton field $\chi$ as relevant degrees of freedom
in addition to the pion field $\pi$,  and  (\ref{lag-an}) is the
anomalous part of the Lagrangian known as Wess-Zumino term. Unlike
HLS theory which has manifest local gauge invariance, (\ref{lag-n})
is gauge fixed to unitary gauge with the density-independent
parameters fixed at a given scale. As such, it lacks the intrinsic
density dependence present in HLS theory with the VM fixed point.
Now what the dilaton does is to mimic this intrinsic dependence by
multiplying the parameters with the expectation value $\la\chi\ra^*$
of the dilaton field  with respect to the vacuum modified by the
density. The dilaton $\chi$ that figures here is to capture the
physics of the spontaneous breaking of scale invariance by the QCD
vacuum. There is a subtlety in this matter in that scale symmetry is
broken both explicitly by the trace anomaly of QCD and spontaneously
by the vacuum. Unlike internal symmetries which are broken both
explicitly and spontaneously,e.g., chiral symmetry, however, scale
symmetry can be broken spontaneously {\it only if} it is broken
explicitly~\cite{freund-nambu}. This makes its implementation a
delicate task. This subtlety can be easily managed for the normal
component of the Lagrangian, and the resulting Lagrangian is
(\ref{lag-n}) using the method of \cite{ellis}.

It is the anomalous part (\ref{lag-an}) -- referred to as
``Wess-Zumino term" --  that poses the problem, both conceptually
and in practice. In general, there are three independent terms
(excluding external fields) in the Wess-Zumino term  with the
coefficients undetermined by the symmetry. The simple form given in
(\ref{lag-an}) is arrived at by choosing the arbitrary coefficients
in a special, ad hoc,  way and by using the equation of motion for
the $\rho$ field~\cite{meissner}. Whether or not this is a
reasonable procedure will be discussed below. What is important for
our discussion is that as it stands, it is scale-invariant and hence
does not couple to $\chi$. To see that this is the main culprit of
the problem we have in this model, consider the energy per baryon
contributed by this term~\cite{vector}:
 \be
 \left(\frac{E}{B}\right)_{WZ}=\frac 14(\frac{3g}{2})^2\int_{Box}
 d^3x\int d^3x^\prime B_0 (\vec{x})
 \frac{{\rm exp}(-m_\omega^*|\vec{x}-\vec{x}^\prime|)}
 {4\pi|\vec{x}-\vec{x}^\prime|} B_0 (\vec{x}^\prime)\label{wzenergy}
 \ee
where ``Box" corresponds to a single FCC cell. This quantity
diverges unless it is screened by $m_\omega^*$. Therefore if the
$\omega$ mass were to go down as required by the vector
manifestation, the skyrmion-skyrmion interactions would become
strongly repulsive with increasing density. This forces the $\omega$
mass $m_\omega^*$, and hence $\la\chi\ra^*$, to increase.  Note that
in HLS theory with the vector manifestation, $g^2$ in
(\ref{wzenergy}) drops to zero as the density approaches the
critical, so the problem is avoided.

The upshot is that although symmetry consideration requires that
there be change-over from skyrmions to half-skyrmions at some high
density~\cite{half-skyrmions,byp1,byp2}, there is a strong repulsion
in the interactions which would make the putative phase transition
occur at very high density. If this were a reality, then there would
be an important ramification on the structure of compact stars. For
instance, such a strong repulsion would rule out possible kaon
condensation at density low enough to be relevant for compact
stars~\cite{pandha-kaon}, increase the maximum stable neutron star
mass above 2 times the solar mass~\cite{akmal} and hence prevent the
formation of low-mass black holes in the
Universe~\cite{BLR-strangeness}. If the HLS scenario with the vector
manifestation is correct, then such a repulsion will be suppressed
giving a totally different picture. At present, the situation is
unclear and which scenario is the viable one remains to be seen.

\section{A Resolution}
Assuming that there is nothing wrong with (\ref{lag-n}), we focus on
the Wess-Zumino term in the Lagrangian. Our objective is to find an
alternative to (\ref{lag-an}) that leads to a behavior consistent
with the VM of HLS theory.

As mentioned, (\ref{lag-an}) has no justification other than its
simplicity~\footnote{The $\omega\cdot B$ term of the form
(\ref{lag-an}) was used by Adkins and Nappi to stabilize the Skyrme
soliton~\cite{adkins-nappi}. However in the presence of a large
number of vector mesons as in holographic dual QCD~\cite{sakai},
this coupling does not exist, and the $\omega$ plays no role in the
stabilization.} For instance, the arbitrary coefficients generally
present in the Wess-Zumno terms are adjusted to give the single term
with the constant fixed so as to give a correct decay rate for
$\omega\rightarrow 3\pi$. However there are compelling indications
that the $\omega\rightarrow 3\pi$ decay does not go through the
direct coupling. Indeed, both HLS theory of \cite{HY:HLS} and
holographic dual QCD that incorporates an infinite tower of vector
mesons~\cite{sakai} show that the decay is totally vector-dominated,
i.e. $\omega\rightarrow\rho\pi\rightarrow 3\pi$. This means that  a
term of the form (\ref{lag-an}) will not figure in $\L_{an}$.  It
should be replaced by terms of the form $\sim {\rm Tr}(\pi dA dA)$
with $A$ the gauge field. It is possible that such terms could
change the interaction structure. Also the way the topological
baryon current $B_\mu$ is derived \`a la
anomaly~\cite{goldstone-wilczek,HHH} suggests that implementing the
trace anomaly of QCD in hidden local symmetric theory is perhaps
subtler than what is presently known.

Various alternative ways-out explored so far~\cite{park}, however,
do not offer a clear-cut solution. Here in the absence of any
reliable clue, we try the simplest, admittedly {\it ad hoc},
modification of the Lagrangian (\ref{lag-an}) that allows a
reasonable and appealing way-out.  Given our ignorance as to how
spontaneously broken scale invariance manifests in matter,  we shall
simply forego the requirement that the anomalous term be scale
invariant and multiply the $\omega\cdot B$ term by $(\chi/f_\chi)^n$
for $n\gsim 2$. We have verified that it matters little whether we
pick $n=2$ or $n=3$~\cite{park}. We therefore take $n=3$: \be
\L_{an}^\prime = \textstyle\frac{3}{2} g (\chi/f_\chi)^3 \omega_\mu
B^\mu \label{lag-anp}\ee
This additional factor has two virtues:

\begin{itemize}
\item [ i)] It leaves meson dynamics in free space
(i.e. $\chi/f_\chi = 1$) unaffected,
since chiral symmetry is realized \`a la sigma model as
required by QCD.

\item [ii)] It plays the role of an effective density-dependent
coupling constant so that at high density, when scale symmetry is
restored and $\chi/f_\chi \rightarrow 0$,  there will be no coupling
between the $\omega$ and the baryon density as required by hidden local
symmetry with the vector manfiestation.

\end{itemize}
The properties of this Lagrangian for the meson ($B=0$) sector are
the same as in our old description. The parameters of the Lagrangian
are determined by meson physics and given in Table 1 in
\cite{vector}.

\section{The B=1 Skyrmion : Hedgehog Ansatz}

The solitons of this effective theory are skyrmions. From
Eq.(\ref{lag}),  the spherically symmetric hedgehog Ansatz for the
$B=1$ soliton solution of the standard Skyrme model can be
generalized to
\begin{equation}
U^{B=1} = \exp(i\vec{\tau}\cdot\hat{r} F(r)),
\label{UBeq1}\end{equation}
\begin{equation}
\rho^{a,B=1}_{\mu=i} = \varepsilon^{ika}\hat{r}^k \frac{G(r)}{gr},
\hskip 2em \rho^{a,B=1}_{\mu=0} = 0, \label{RBeq1}\end{equation}
\begin{equation}
\omega^{B=1}_{\mu=i} = 0, \hskip 2em \omega^{B=1}_{\mu=0} = f_\pi
W(r), \label{WBeq1}\end{equation}
and
\begin{equation}
\chi^{B=1} = f_\chi C(r).
 \label{CBeq1}\end{equation}
The boundary conditions that the profile functions satisfy at
infinity are
 \begin{equation}
F(\infty)=G(\infty)=W(\infty)=0, \hskip 2em C(\infty)=1
 \end{equation}
while near the origin,
\begin{equation}
F(0) = \pi, \hskip 2em G(0) = -2, \hskip 2em W^\prime(0)=0,
\hskip 2em C^\prime(0)=0.
\end{equation}

The equations of motion for the various profile functions $F(r)$,
$G(r)$, $W(r)$ and $C(r)$ follow from the minimization of the soliton
mass. The numerical results on the properties of the $B=1$ hedgehog
skyrmion, the mean-square baryon number radius, the mean-square
energy radius, the mass and the contributions to the mass from the
different meson terms are reproduced in Table 1 for different values
of the dilaton mass.

\begin{table}
\caption{Baryonic mean square radius, energy mean square radius and
mass for the skyrmion. We also show the different contributions to
the mass from the various terms.}
\begin{center}
\begin{tabular}{ccccccccccc}
\hline $m_\chi$ & $\sqrt{\langle r^2\rangle_B}$ & $\sqrt{\langle
r^2\rangle_E}$ & $M_{sol}$ & $E_\pi$ & $E_{m_\pi}$ & $E_{\pi\rho}$ &
$E_\rho$  & $E_\omega$ & $E_{WZ}$ & $E_\chi$
\\
\hline
720   & 0.12 & 0.39 &  916 &  27 &  0 & 178 & 503 & -1 & 1 & 206 \\
1000  & 0.12 & 0.37 &  956 &  35 &  1 & 184 & 524 & -1 & 2 & 211 \\
1200  & 0.13 & 0.37 &  982 &  53 &  3 & 177 & 526 & -1 & 3 & 222 \\
1300  & 0.49 & 0.59 & 1410 & 719 & 42 & 34 & 375 & -238 & 474 & 5 \\
1500  & 0.49 & 0.59 & 1439 & 748 & 42 & 33 & 370 & -246 & 489 & 2 \\
2000  & 0.50 & 0.60 & 1459 & 757 & 44 & 32 & 367 & -248 & 496 & 1 \\
\hline
\end{tabular}
\end{center}
\end{table}

Two regimes can be distinguished in the table as a function of
dilaton mass chosen in a wide range. Let us call them  the
long-range dilaton (LRD) regime, which occurs for ``small" dilaton
mass, say, $m_\chi \leq 1200$ MeV, and the short-range dilaton (SRD)
regime, which occurs for ``large" mass, say. $m_\chi\geq 1300$ MeV.
They show very different properties which we describe, and whose
origin we shall discuss.

In the LRD regime, the baryon density is concentrated in a small
volume ($\sqrt{ \langle r^2\rangle_B} < 0.15 $ fm), and  the
skyrmion is light ($M_{sol} < 1000 $ MeV). However, note that the
radius associated with the energy of the skyrmion is appreciable
($\sqrt{\langle r^2 \rangle_E} \sim 0.4 $fm). We can understand
these two sizes as follows. Although the dilaton squeezes the
skyrmion, so the baryon number gets concentrated in a small core, a
large $\rho$ meson cloud, carrying most of the energy, surrounds
this skyrmion core. The $\omega$ meson, the Wess-Zumino ($\omega-B$)
term and the $\pi$ mass terms contribute negligily to the mass. Even
the pion dynamic term is small. Most of the mass comes, therefore,
from the $\rho$ couplings and the $dilaton$ terms.

Figure 1 (left) in which the profile function of all fields are
shown proves to be very illuminating for understanding what is going
on. The profile function of the dilaton is particularly clarifying.
For large values of the skyrmion radius,  the dilaton field is as in
free space, but for small values, it drops to zero. Thus we see that
scale symmetry is realized differently in the interior than in the
exterior of the skyrmion. While in the exterior it is spontaneously
broken,  it is partially restored in the interior.  If we look at
the profiles of other mesons, we see that those of $\pi$ and
$\omega$ are large,  changing only in the small interior region,
while the $\rho$ meson profile extends further out, confirming our
previous discussion. Since both scale symmetry and chiral symmetry
are partially restored in the inside region, the pion mass term and
the pion dynamic term become also small (i.e. Wigner mode). Finally
the $\omega\cdot B$ coupling and the $\omega$ term also become
small, perhaps reflecting that the vector
manifestation~\cite{HY:HLS} is effective. The long-range dilaton
makes the skyrmion size small and the baryon number density large.
Thus a sort of phase transition takes place to a symmetry restored
phase in its interior with vanishing $\omega$-couplings, resembling
the chiral bag model. Note that most of the mass comes from the
$\rho$ and the dilaton. The $\rho$ is not squeezed, so the energy
radius is large due to its cloud. Thus, this regime is dominated by
vector mesons ($\rho$) and partial restored scale symmetry, the
dilaton.

We find  a completely different structure in the SRD regime. Here
the skyrmion is large ($\sqrt{\langle r^2\rangle_B}  \sim  0.5 $
fm), with the baryon density radius nearly coinciding with the
energy radius ($\sqrt{\langle r^2\rangle_B}  \sim  0.6 $ fm). It is
also heavy ($M_{sol} \sim 1450 $ MeV). The contribution to the mass
from the dilaton comes out to be very small, while the pion
contribution from the dynamic term is huge. The mass term gives a
sizeable contribution. The contribution from the $\omega$ is large
and attractive, and that from the $\omega\cdot  B$ term even larger
and repulsive. The $\rho$ contribution is sizeable, but smaller than
in the LRD regime, with its cloud less important.  Figure 1 (right
panel) shows that the dilaton profile is practically constant and
non-zero, indicating that chiral symmetry is realized in Goldstone
mode. Here the pion and the $\omega$ are indispensable for the
description of the strong force.

What is remarkable in the numerical results is that the transition
between the two regimes is abrupt. This is easy to understand
mathematically. For a fixed $f_{\chi}$, the dilaton potential
changes sign at a particular value of $m_\chi$ which induces an
abrupt change from Goldstone to Wigner.

To summarize, we note that with the scale factor multiplied to the
Wess-Zumino term, the role of the $\omega$ field is tightly locked
to the behavior of the dilaton field that controls the realization
of the two regimes appearing in the single skyrmion system.

\begin{figure}[tbp]
\centerline{\epsfig{file=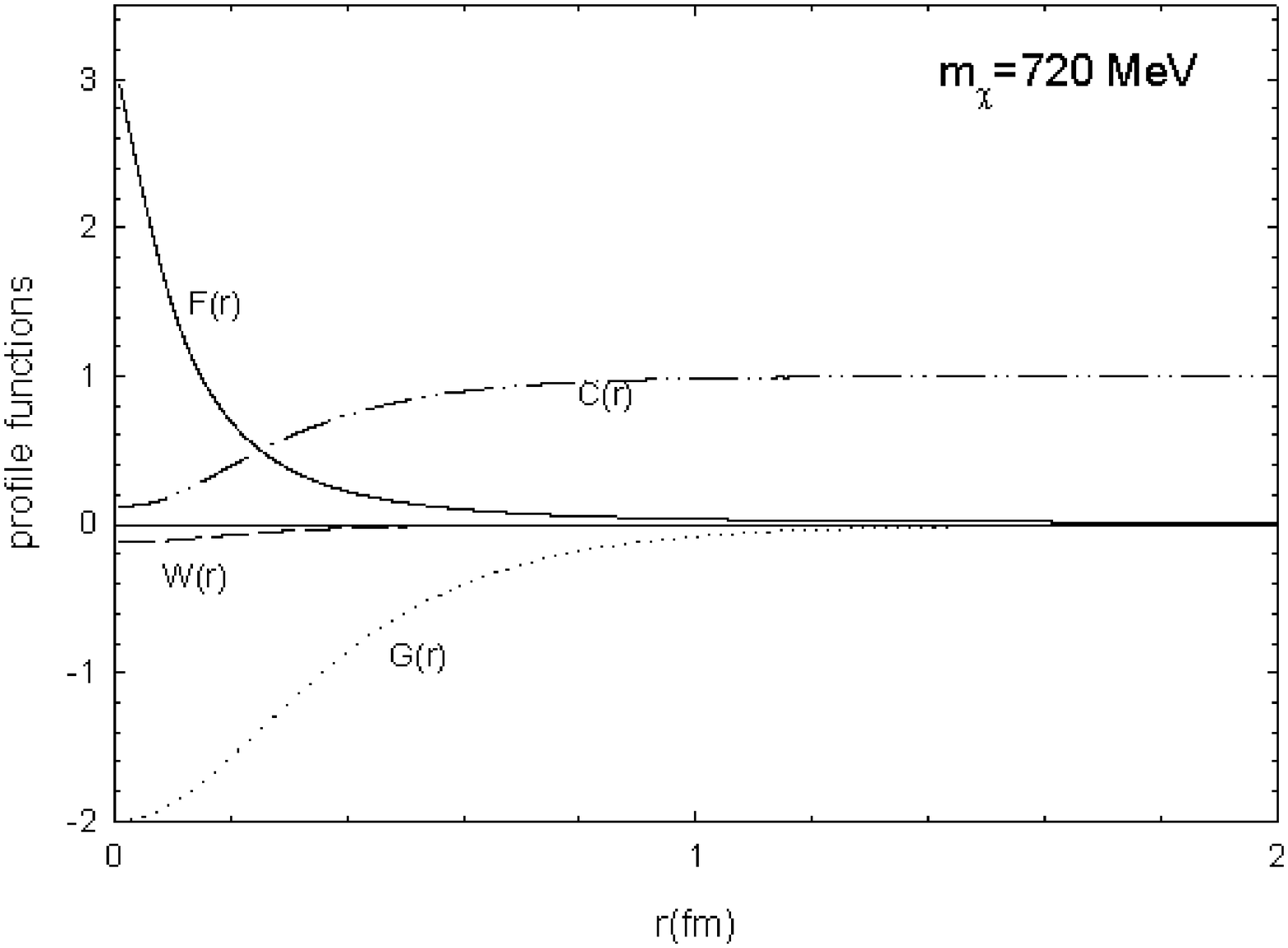,width=8cm,angle=0}
\epsfig{file=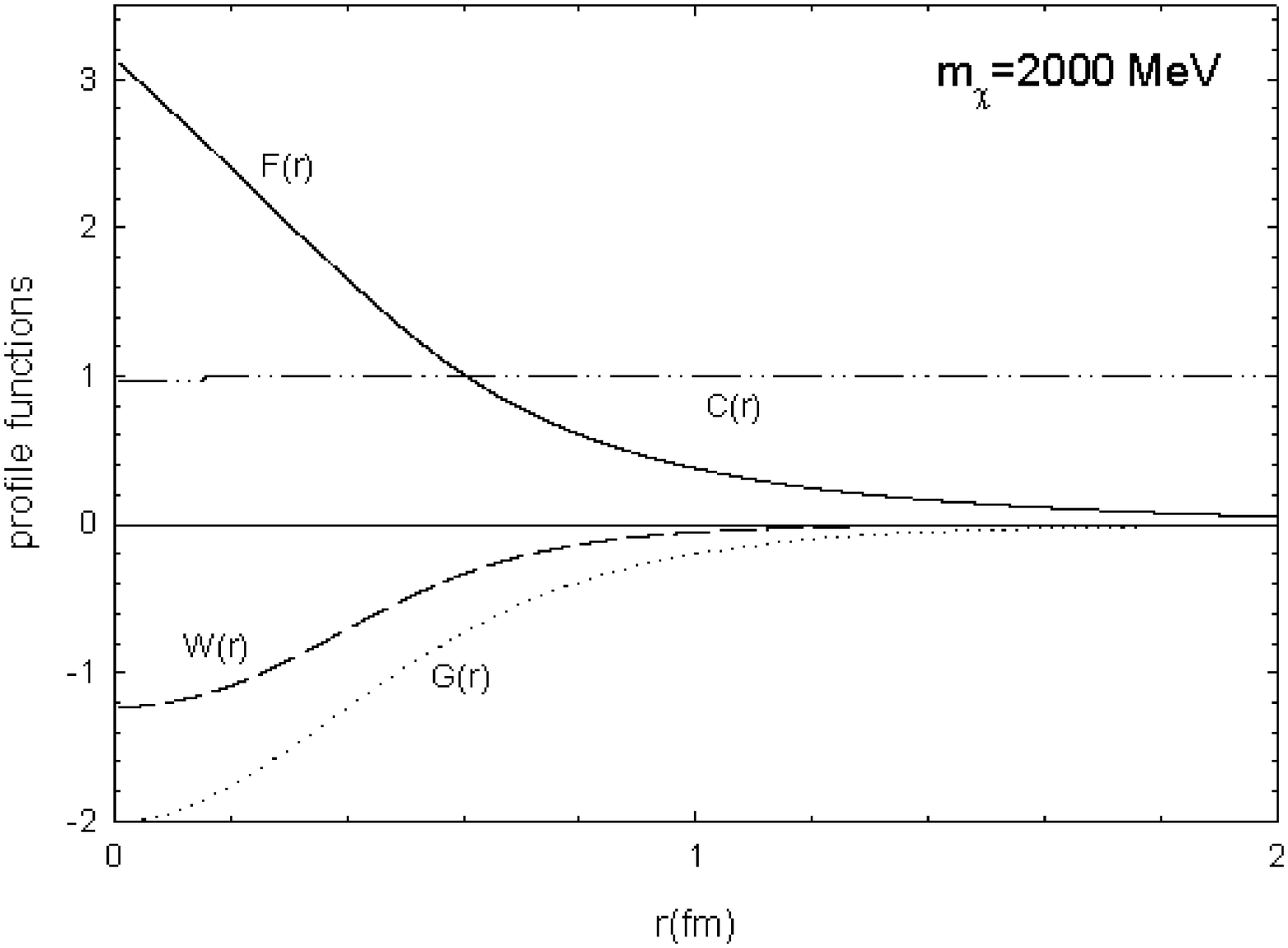,width=8cm,angle=0}} \caption{Small skyrmion
solution obtained with $m_\chi=720$ MeV and large skyrmion with
$m_\chi=2000$ MeV.}
\end{figure}

\section{Phase Transitions in Dense Skyrmion Matter}

The procedure to describe skyrmion matter follows closely what was
done in \cite{vector}. It suffices to multiply the scaling factor to
the $\omega - B$ coupling term. At low densities, skyrmion matter is
described  by an FCC crystal where the nearest neighbor interactions
are arranged to have attractive relative orientations. In order to
exemplify the phase transitions,  we consider the $m_\pi = 0 $ case
for which the effects are more dramatic.\footnote{In Table 1, a
non-vanishing pion mass is put in as a cut-off to regularize
divergences in evaluating the rms radii, $\sqrt{\langle
r^2\rangle_B}$ and $\sqrt{\langle r^2\rangle_B}$. }

We show in Fig. 2 the numerical results on $\langle\chi\rangle$ and
$\langle\sigma\rangle$ for an exemplary  ``low" dilaton mass and a
``high" dilaton mass as a function of the FCC parameter $L_F$, which
is related to the baryon density by $\rho_B = 1/2L_F^3$
\footnote{Nuclear matter density $\rho_0 = .17/fm^3$ corresponds to
$L_F \sim  1.43$ fm.}. As in the single skyrmion case discussed in
the previous section,  there are again two regimes,  the LRD regime
shown on the left panel of Fig. 2 and the SRD regime shown on the
right panel.
\begin{itemize}
\item In the LSD regime, the symmetry-favored half-skyrmion
phase with $f_\pi\propto \la\chi\ra^*\neq 0$ and $\la\sigma\ra\propto \la\bar{q}q\ra=0$
shrinks and the phase transition is characterized
by the vanishing $\langle\chi\rangle$. The phase transition occurs at about  three times
the nuclear matter density $  n_0$, for the set parameters given \cite{vector}.
Up to but below the critical point, say, $n=n_c-\epsilon$ where $n_c$ is the
critical density,  the skyrmion matter is of an FCC crystal. One cannot say
what it is precisely at $n=n_c$, but symmetry consideration suggests that it
could be in the half-skyrmion phase.
\item The situation is completely different in the SRD regime.
Here the Goldstone phase in an FCC crystal changes over at $n=n_p <n_c$ to a
half-skyrmion phase in BCC crystal with $f_\pi\propto \la\chi\ra^*\neq 0$
and $\la\sigma\ra=0$ and then to the Wigner phase with $f_\pi=\sigma=0$.
As discussed in \cite{MR-halfskyrmion}, when interpreted in HLS theory,
the half-skyrmion phase  corresponds to the phase in which $f_\pi=f_\sigma\neq 0$
(where the subscript $\sigma$ represents the longitudinal component of
the $\rho$ meson) and $g\simeq 0$.  When $g=0$ with $f_\pi=f_\sigma\neq 0$,
there is an enhancement of symmetry $SU(2)^2\rightarrow SU(2)^4$ as pointed
out by Georgi~\cite{georgi}.
    \end{itemize}

\begin{figure}[tbp]
\centerline{\epsfig{file=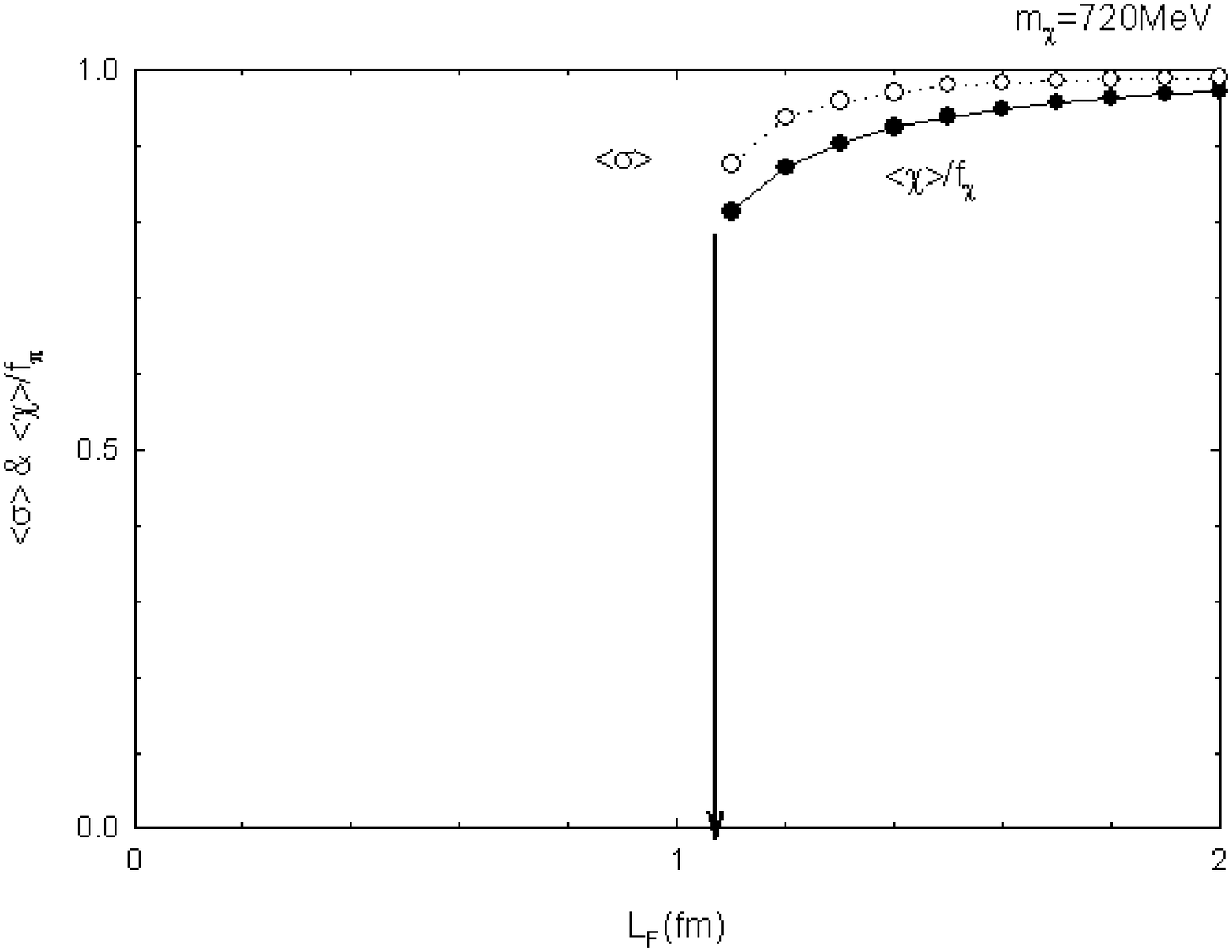,width=8cm,angle=0}
\epsfig{file=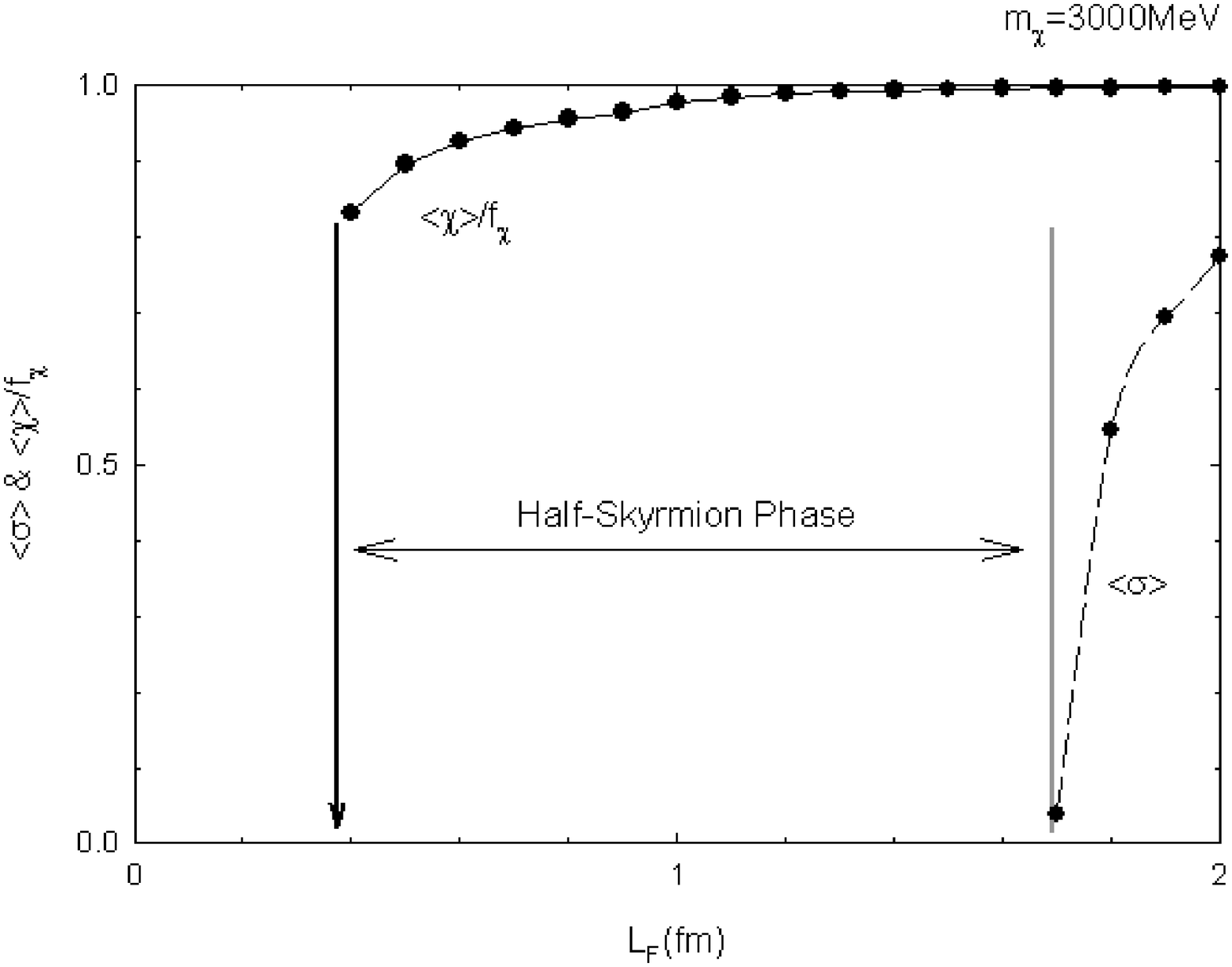,width=8cm,angle=0}} \caption{Behavior of
$\langle\chi\rangle$ and $\langle\sigma\rangle$ as a funcion
of lattice size $L_F$ in the
LRD regime
 for $m_\chi=720$ MeV (left figure) and in
the SRD regime for $m_\chi=3000$ MeV (right figure).}
\end{figure}

We show in Fig. 3 the phase diagram  as a function of $L_F$ and
dilaton mass. This diagram illustrates an intricate nature of the
phase structure. As the dilaton mass increases, the critical density
at which $\langle \chi\rangle^*=0$  sharply increases to a very
large value: This occurs before $\langle \sigma\rangle$ vanishes.
For some values of the dilaton mass, we observe an abrupt change
leading to a different phase transition scenario in which  $\langle
\sigma \rangle$ vanishes at low density, giving rise to a pseudogap
phase~\cite{byp1,byp2}, which remains until $\langle \chi\rangle$
vanishes at high density.

\begin{figure}[tbp]
\centerline{\epsfig{file=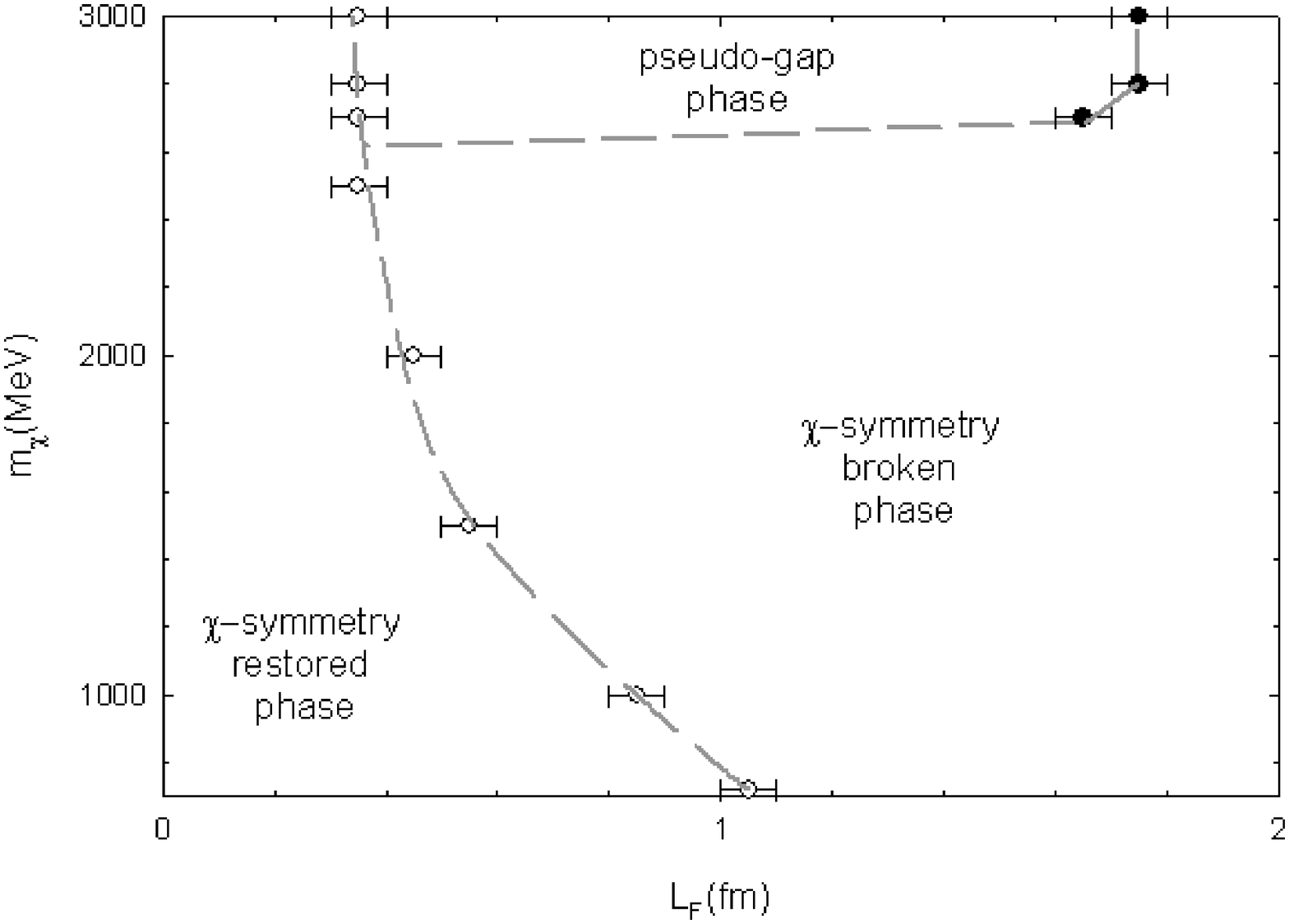,width=10cm,angle=0}}
\caption{Phase diagram as a function of lattice size $L_F$ and
dilaton mass $m_\chi$.}
\end{figure}

Since we have changed only the $\omega\cdot B$ coupling term, the
scaling behavior of the in-medium physical quantities remains the
same as those in \cite{vector}, which emphasize the importance of
the quantities shown in Fig.2.
\section{Concluding Remarks}
In our effort to find a unified approach to dense matter, i.e.
matter at densities higher than that of the normal nuclear matter,
which has remained ill-understood up to date, we have uncovered a
hitherto unsuspected role that the $\omega$ meson and the dilaton --
associated with the trace anomaly of QCD -- play in the structure of
dense skyrmion matter. A straightforward implementation of the
dilaton field to a gauged skyrmion Lagrangian exposed a serious
difficulty in the description of the properties of dense skyrmion
matter.  We identified the source of the problem in an ill behavior
of the $\omega$ meson in nuclear matter, due primarily to its
coupling to the baryon current in the Wess-Zumino term in the
effective Lagrangian. In this article, we present a simple and
elegant solution to the problem, which exposes a remarkable
interplay between the light-quark vector mesons of hidden local
symmetry and the dilaton of scale symmetry. Given the poorly
understood intricacy involved in the way the dilaton figures in the
scale-symmetry breaking, our solution does not yet receive a
justification from QCD. An effort is being made to find -- if any --
such a connection to QCD~\cite{park}.

As in our previous work without vector mesons, the skyrmion matter
possesses two phases: one which is described by an FCC crystal and
another which is described by a half-skyrmion BCC crystal. The main
result of the present investigation is the discovery that the
dilaton wholly governs the phase transition. We have seen that two
regimes are present: A first in which the dilaton dynamics is
long-ranged and which leads to what may correspond to a conventional
chiral phase transition in which the spontaneously broken scale
invariance and chiral symmetry are restored simultaneously;  a
second in which the dilaton dynamics is short-ranged and which leads
to a two-step transition. Initially, chiral symmetry changes at
relatively large densities from a spontaneously broken phase to a
pseudo-gap phase, and ultimately scale invariance is restored and
the system goes into a symmetric phase. The change from one regime
to the other is extremely sensitive to a particular value of the
dilaton mass, which is certainly dependent on the meson parameters
used, and which is associated with a change in sign in the potential
of the dilaton which moves abruptly from a double well scenario to a
single well scenario. The dilaton mass which delineates the two
different scenarios is fairly high, a few times the chiral scale.
Nature is most likely in the LRD, so the model favors the phase
transition without going through the half-skyrmion phase (or
equivalently Georgi's vector symmetry phase~\cite{georgi}), but
since we do not know what the correct structure of the Wess-Zumino
term in medium is with respect to scale symmetry, it remains to be
seen whether this is what Nature adopts. Even so, what is truly
remarkable is that a mild change of the scale structure of the
Wess-Zumino term makes such a dramatic change in the properties of
dense matter. In particular, the effect of the dilaton mass
depending on whether it is of the order of the vector meson mass or
several times the vector meson mass is quite striking and
unintuitive.

It is interesting to note that the same mechanism operates quite
similarly in the single skyrmion description. Thus in the SRD
regime,  the skyrmion is squeezed to a small size, while the full
size of the skyrmion is obtained by a large $\rho$ meson cloud,
while in the LRD regime,  the skyrmion is large, and the $\rho$
cloud  smaller. However the mass at which we move from one regime to
another is different for the single skyrmion and  the skyrmion
matter. This can be simply understood  by that the potential for the
latter is nonperturbatively dressed by the medium as shown in
\cite{byp1}. Finally we see that in the SRD regime,  the pion and
$\omega$-meson  fields play insignificant role, while they are of
fundamental importance in the LRD regime. The finding here could
offer a way to understand how confinement works in chiral models.

\section*{Acknowledgments}

Vicente Vento is grateful for the hospitality extended to him by the
Physics Department of Chungnam National University and the Physics
Department, TH-Unit, at CERN. VV acknowledges support from the
agreement CSIC-KOSEF. This work was partially supported by grants
MEC-FPA2007-65748-C02-01, MEC-Movilidad-PR2007-0048 and
GV-GRUPOS03/094 (VV) and research fund of CNU in 2005(BYP).

\end{document}